\documentclass{PoS}
\usepackage{amsmath,amssymb,latexsym, cancel}
\usepackage{amssymb,latexsym}
\usepackage{graphicx}
\usepackage{float}

\title{AdS/QCD Modified Soft Wall Model and Light Meson Spectra}

\ShortTitle{AdS/QCD Modified Soft Wall Model and Light Meson Spectra}

\author{\speaker{Santiago Cort\'es} \\
        Departamento de F\'{\i}sica, Univ. de Los  Andes, 111711 Bogot\'a, Colombia.\\
        E-mail: \email{js.cortes125@uniandes.edu.co}}

\author{Miguel \'Angel Mart\'{\i}n Contreras\\
        Departamento de F\'{\i}sica, Univ. de Los  Andes, 111711 Bogot\'a, Colombia. \\
				Departamento de Ciencias B\'asicas, Univ. Cat\'olica de Colombia, 111311, Bogot\'a, 			Colombia. \\
        E-mail: \email{ma.martin41@uniandes.edu.co}}
				
\author{Jos\'e Rolando Rold\'an\\
        Departamento de F\'{\i}sica, Univ. de Los  Andes, 111711 Bogot\'a, Colombia.\\
        E-mail: \email{jroldan@uniandes.edu.co}}

\abstract{We analyze here the mass spectrum of light vector and scalar mesons applying a novel approach where a modified soft wall model that includes a UV-cutoff at a finite $z$-position in the AdS space is used, thus introducing an extra energy scale. For this model, we found that the masses for the scalar and vector spectra are well fitted within $\delta_{RMS}=7.64\%$ for these states, with non-linear trajectories given by two common parameters, the UV locus $z_{0}$ and the quadratic dilaton profile slope $\kappa$. We also conclude that in this model, the $f_{0}(500)$ scalar resonance cannot be fitted holographycally as a $q\overline{q}$ state since we cannot find a trajectory that include this pole. This result is in agreement with the most recent phenomenological and theoretical methods.}

\FullConference{XVII International Conference on Hadron Spectroscopy and Structure - Hadron2017\\
		25-29 September, 2017\\
		University of Salamanca, Salamanca, Spain}

\begin{document}

\section{Introduction and Formalism}

The AdS/CFT correspondence \cite{Maldacena:1997re} has been largely used to describe nonperturbative QCD-like phenomena which are unreachable by regular QFT methods. One example is given by the dynamics of the lightest pseudoscalar mesons, which is properly described via the Effective Field Theory approach of Chiral Perturbation Theory (ChPT) \cite{Gasser:1983yg}. If resonant states are to be included, a proper unitarization method has to be considered.

In this case, we will use a bottom-up approach known as the AdS/QCD Soft-Wall model whose Lagrangian reads \cite{Karch:2006pv}

\begin{align}
I=&-\frac{1}{2\,g_S^2}\int{d^5x\,\sqrt{-g}\,\exp[-\Phi\left(z\right)]\,\left[\partial_n\,S\,\partial^n\,S+m_5^2\,S^2\right]}\notag \\
&-\frac{1}{4\,g_V^2}\,\int{d^5x\,\sqrt{-g}\,\exp[-\Phi\left(z\right)]F_{mn}\,F^{mn}}, \label{general-action}
\end{align}

\noindent where $S\left(z,x^\mu\right)$ is a massive scalar field dual to the scalar mesons and $F_{mn}=\partial_m\,A_n-\partial_n\,A_m$ is given in terms of the massless abelian gauge field $A_m\left(z,x^\mu\right)$. The constants $g_S$ and $g_V$ fix the units of the action in terms of the number of colors $N_c$ as usual. As it can be seen, chiral symmetry breaking effects are not taken into account.

The geometric background that explicitly breaks the conformal invariance is given by the sliced AdS Poincare patch \cite{Braga:2015jca}

\begin{equation}\label{geometry}
dS^2=\Theta\left(z-z_0\right)\,\frac{R^2}{z^2}\left[dz^2+\eta_{\mu\nu}\,dx^\mu\,dx^\nu\right],
\end{equation}

\noindent with $\Theta\left(z\right)$ the Heaviside step function that gives the UV D-brane (D-Wall) locus. The Minkowski metric has the signature $(-, +, +, +)$. All of this will allow us to define the mass spectrum of light scalar and vector mesons as functions of two energy scales, namely, the D-wall locus $z_0$ and the dilaton constant $\kappa$, as showed in \cite{Braga:2015jca}.

\section{Soft-Wall model Light Meson Spectra}

%\subsection{Vector Mesons}

We begin our analysis by taking the light vector meson action that reads 

\begin{equation}
I_{V}=-\frac{1}{4\,g_V^2}\,\int{d^5x\,\sqrt{-g}\,\exp[-\Phi\left(z\right)]F_{mn}\,F^{mn}}.
\label{Vector-action}
\end{equation}

After taking small variations in $A_{\mu}$ and imposing the gauge condition $A_z=0$, we obtain an On-Shell Boundary action given by

\begin{equation}
I_{\text{V On-Shell}}^{\text{Boundary}}=-\frac{R}{2g_{V}^{2}}\int{d^{4}x\left. \frac{\exp(-\kappa^{2}z^{2})}{z}A_{\mu}\,\partial_{z}\,A^{\mu}\right|_{z_{0}}}.
\label{Vec-osbdaction}
\end{equation}

Two-point functions are straightforwardly obtained after solving the vector equation of motion by introducing Fourier transformed vector fields 

\begin{equation}
A^{\mu}(z,x^{\mu})=\frac{1}{(2\pi)^{4}}\int{d^{4}q\,\exp(-iq_{\mu}x^{\mu})\,\,v_{\mu}(z,q)},
\label{vector-ft}
\end{equation}

\noindent where we write $v_{\mu}(z,q)$ as a function of the source term $v_{\mu}^{0}(q)$ and the Bulk-to-Boundary propagator $V(z,q)$ as $v_{\mu}(z,q)=v_{\mu}^{0}\left(q\right)\,V(z,q)$ Hence, we obtain that $V(z,q)$ holds with the following equation of motion:

\begin{equation}
\partial_{z}\left[\frac{\exp(-\kappa^{2}z^{2})}{z}\,\partial_{z}V(z,q)\right]+\frac{q^{2}}{z}\,\exp(-\kappa^{2}z^{2})V(z,q)=0.
\label{eom-BtoBp}
\end{equation}

A proper vector mass spectrum is obtained from \ref{Vec-osbdaction} through its two-point function. In order to obtain it, we have to consider the Fourier transformation of the vector fields so that a Bulk-to-Boundary propagator $V(z,q)$ is properly introduced, along with a source term $v_{\mu}^{0}(q)$. We check that the solution of $V(z,q)$ yields with the following result:

\begin{equation}
V(z,q)=c_{1}\,\kappa^{2}\,z^{2}\,\text{}_{\,1}F_{1}\left(1-\frac{q^{2}}{4\kappa^{2}},2,\kappa^{2}z^{2}\right),
\label{vec-regsol}
\end{equation}

\noindent where $\text{}_{\,1}F_{1}(1-q^{2}/4\kappa^{2},2,\kappa^{2}z^{2})$ is the Kummer confluent hypergeometric function and $c_{1}$ is a normalization constant. Since the vector two-point function $G^{\mu\nu}(q^{2})$ has to hold with $G^{\mu\nu}(q^{2})=\eta^{\mu\nu}\,\Pi(q^{2})$, we obtain after normalizing (\ref{vec-regsol}) such that $V(z_{0})=1$, the following relation for $\Pi(q^{2})$:

\begin{equation}
\Pi(q^{2})=-\frac{R\,\exp(-\kappa^{2}z_{0}^{2})}{g_{V}^{2}z_{0}^{2}}\left[\frac{2}{z_{0}}+\kappa^{2}z_{0}\left(1-\frac{q^{2}}{4\kappa^{2}}\right)\frac{_{1}F_{1}\left(2-\frac{q^{2}}{4\kappa^{2}},3,\kappa^{2}z_{0}^{2}\right)}{_{1}F_{1}\left(1-\frac{q^{2}}{4\kappa^{2}},2,\kappa^{2}z_{0}^{2}\right)}\right].
\label{pi-factor2}
\end{equation} 

In order to obtain the scalar meson sector, we follow a similar procedure to find a two-point function from the scalar action

\begin{equation}
I_{S}=-\frac{1}{2g_{S}^{2}}\int{d^{5}x\,\sqrt{-g}\,\exp[-\Phi\left(z\right)]\,\left[\partial_n\,S\,\partial^n\,S+m_5^2\,S^2\right]},
\label{scalar-action}
\end{equation}

\noindent whose associated equation of motion and solution for the Bulk-to-Boundary propagator respectively read

\begin{equation}
\partial_{z}\left[\frac{\exp(-\kappa^{2}z^{2})}{z^{3}}\partial_{z}\overline{v}(z,q)\right]+\frac{\exp(-\kappa^{2}z^{2})}{z^{3}}q^{2}\overline{v}(z,q)+\frac{3\exp(-\kappa^{2}z^{2})}{z^{5}}\overline{v}(z,q)=0,
\label{scalar-eom-vz}
\end{equation}

\begin{equation}
\overline{v}(z,q)=\overline{c}_{1}\,\kappa^{3}z^{3}\,\text{}_{1}F_{1}\left(\frac{3}{2}-\frac{q^{2}}{4\kappa^{2}},2,\kappa^{2}z^{2}\right).
\label{scalar-propbtb}
\end{equation}

We obtain the latter relations after writing the Fourier-transformed scalar field as $S(z,q)=S^{0}(q)\overline{v}(z,q)$. Its respective normalized two-point function is such that 

\begin{equation}
\Pi_{S}(q^{2})=-\frac{R^{3}}{g_{S}^{2}}\frac{\exp(-\kappa^{2}z_{0}^{2})}{z_{0}^{\,3}}\left[\frac{3}{z_{0}}+\kappa^{2}z_{0}\left(\frac{3}{2}-\frac{q^{2}}{4\kappa^{2}}\right)\frac{_{1}F_{1}\left(\frac{5}{2}-\frac{q^{2}}{4\kappa^{2}},3,\kappa^{2}z_{0}^{2}\right)}{_{1}F_{1}\left(\frac{3}{2}-\frac{q^{2}}{4\kappa^{2}},2,\kappa^{2}z_{0}^{2}\right)}\right].
\label{scalar-twop2}
\end{equation}

Our theoretical predictions \cite{Cortes:2017lgz} are obtained from the poles of (\ref{pi-factor2}) and (\ref{scalar-twop2}) after adjusting the dilaton parameters $z_{0}$ and $\kappa$ to $z_{0}=\text{5 GeV}^{-1},\,
\kappa=\text{0.45 GeV}$. We compare these results with the most recent PDG data \cite{Olive:2016xmw} (as showed in Tables \ref{tab:vecmes} and \ref{tab:scalmes}), thus obtaining a RMS error $\delta_{\text{RMS}}=\text{7.64}\%$.

\begin{table}[H]
\centering
\begin{tabular}{ccccccc}   \hline \hline
	$\rho$ & & $M_{\text{th}}$ (GeV)& & $M_{\text{exp}}$ (GeV) & & $\% M$  \\ \hline
	$\rho(775)$ & & 0.975  & & 0.775 & & 20.53  \\ 
	$\rho(1450)$  & & 1.455 & & 1.465  & & 0.66 \\
	$\rho(1570)$ & & 1.652 & & 1.570 & & 4.96 \\
	$\rho(1700)$ & & 1.829 & & 1.720 & & 5.97 \\ 
	$\rho(1900)$ & & 1.992 & & 1.909 & & 4.15 \\
	$\rho(2150)$ & & 2.142 & & 2.153 & & 0.50   \\ \hline \hline
\end{tabular}
\caption{Mass spectrum for $\rho$ vector mesons with $\kappa=0.45$ GeV and $z_0=5$ GeV$^{-1}$.}
\label{tab:vecmes}
\end{table}

\vspace{-0.46cm}

\begin{table}[H]
\centering
\begin{tabular}{ccccccc}   \hline \hline
	$f_{0}$ & & $M_{\text{th}}$ (GeV)& & $M_{\text{exp}}$ (GeV) & & $\% M$  \\ \hline
	$f_{0}(980)$ & & 1.070  & & 0.99 & & 7.46  \\ 
	$f_{0}(1370)$  & & 1.284 & & 1.370  & &  5.11 \\
	$f_{0}(1500)$ & & 1.487 & & 1.504 & & 1.13 \\
	$f_{0}(1710)$ & & 1.674 & & 1.723 & & 2.93 \\ 
	$f_{0}(2020)$ & & 1.846 & & 1.992 & & 7.94 \\
	$f_{0}(2100)$ & & 2.153 & & 2.101 & & 2.39   \\ 
	$f_{0}(2200)$ & & 2.292 & & 2.189 & & 4.49   \\  
	$f_{0}(2330)$ & & 2.424 & & 2.314 & & 4.52   \\  \hline \hline
\end{tabular}
\caption{Mass spectrum for $f_{0}$ scalar resonances with $\kappa=0.45$ GeV and $z_0=5.0$ GeV$^{-1}$.}
\label{tab:scalmes}
\end{table}

\section{Conclusions} 

We show here that light meson spectra are quite well reproduced after minimizing the amount of parameters of the model; however, the ground state of the scalar sector, i.e., the $f_{0}(500)$ cannot be holographically reproduced as a $q\overline{q}$ state, unlike what happens after introducing quark condensates and masses via chiral symmetry breaking effects, as in \cite{Vega:2010ne,kapusta}.

\section*{Acknowlegdments}

We want to thank Facultad de Ciencias and Vicerrector\'{\i}a de Investigaciones of Universidad de los Andes for financial support.

\end{document}